\documentclass[11pt]{article}
\usepackage{fullpage}
\usepackage{makeidx}
\usepackage{amstext}
\usepackage{amssymb}
\usepackage{amsmath}
\usepackage{amsthm}
\usepackage[cp866]{inputenc}
\usepackage{graphicx}

\begin{document}
\title{The spectra of the oscillating shear flows.}
\author{Sergey Guda\\ Southern federal university\\ Rostov-on-Don, Russia\\ gudasergey@gmail.com}
\date{\today}
\maketitle

\begin{abstract} We study the spectral  problems for  the  spatially  periodic  flows of inviscid incompressible fluid. The basic flows under consideration are the shear flows whose  profiles oscillate on  high frequencies.   For such flows, we present asymptotic expansions of the unstable eigenvalues in the case when the limit spectral problem has multiple eigenvalues.
\end{abstract}

\section*{Introduction} The Euler equations of inviscid incompressible fluid admit   simple steady solutions known as the shear flows. Such solutions (in the Cartesian coordinates) are written in the form
\[
\mathbf{V}=(U(y),0,0)\quad P\equiv const
\]
where $\mathbf{V}$ denotes the velocity field, $P$ denote the pressure, and $U$ is referred to as the flow profile. Generally, function $U$ can be chosen arbitrarily but we restrict our considerations within the case of smooth $2\pi-$periodic profiles.
For such flows, we study 2D spatially periodic  perturbations using the linear approximation. Let  $G$ denote  the stream function for the velocity perturbation;  hence  the instant  velocity of the perturbed flow is
\[
 \mathbf{V}+(G_y(x,y,t),-G_x(x,y,t),0).
\]
Then the  linear equation for perturbations takes the form
\begin{equation}
 \Delta G_t+U(y)\frac{\partial}{\partial x}\Delta G-U''\frac{\partial G}{\partial x}=0,\ \Delta=\partial^2_x+\partial^2_y.
\label{urvozm}
\end{equation}
 Since Eq. (\ref{urvozm}) is invariant with respect to the  the translations both in  $t$ and in $x$, it is natural to seek for
\[
G(x,y,t)=e^{\sigma t}e^{i\alpha x}\Phi(y).
\]
This  substitution leads us to the periodic spectral problem for the Rayleigh equation
\begin{equation}
(\sigma+i\alpha U)\left( \frac{d^2}{d y^2} -\alpha^2 \right)\Phi-i\alpha U''\Phi =0;\quad \Phi(y)=\Phi(y+2\pi).
\label{Rayleigh}
\end{equation}
If there exists an eigenvalue $\sigma$ with $\mathrm{Re}\sigma>0$ then there exists spatially periodic perturbation that grows exponentially in $t$ and the basic shear flow is treated as unstable one. Following to \cite{BelFr} and  \cite{FrVishik}, we look for such instability in the case of rapidly oscillating basic flow.

 Let us consider profile $U(y)=W(s)$, where $W$~-- smooth $2\pi$-periodic in $s$, and  $s=my$ with integer $m$ ($m>>1$).   Let us seek for the solutions of \eqref{Rayleigh} in the form
\begin{equation}
\Phi(y)=e^{iny}f(s),\label{Phin}
\end{equation}
where $n$ is integer and $f$ is unknown $2\pi$-periodic function. Substituting \eqref{Phin} into \eqref{Rayleigh} yields equation for $f$:
\begin{equation}
H(\sigma,\varepsilon)f=0, \label{ishodnaya}
\end{equation}
where operator $H$ is defined by
\begin{equation*}
\begin{split}
&Hf=(\sigma+i\alpha W)f''-i\alpha W''f+\\
+2in\varepsilon(&\sigma+i\alpha W)f'-\varepsilon^2(\alpha^2+n^2)(\sigma+i\alpha W)f,
\end{split}
\end{equation*}
accent means differentiation by variable $s$. We construct  the  asymptotic for the unstable eigenvalues $\sigma$ in the case of $\varepsilon=\frac1m\to 0$.   Our approach follows  to that of \cite{BelFr} and \cite{FrVishik}.   In more details, let us write the problem (\ref{ishodnaya}) in the form
\begin{eqnarray}
&& H_0f+\varepsilon H_1f+\varepsilon^2 H_2 f=0, \label{vved0}\\
&& H_0f=\phi f''-\phi''f;\quad \phi(s)=\sigma+i\alpha W(s); \label{H0}\\
&& H_1f=2in\phi f';\label{H1};\\
&& H_2f=-(\alpha^2+n^2)\phi f;\label{H2}.
\end{eqnarray}
 Assuming that  $\sigma\to\sigma_0$ when $\varepsilon\to 0$ we get
\begin{equation}
H_{00}f\equiv (\sigma_0+i\alpha W(s)) f''-i\alpha W''f=0. \label{vvedH00}
\end{equation}
Eq. \eqref{vvedH00} has periodic solution
\[
f(s)=C(\sigma_0+i\alpha W(s))
\]
for every  $\sigma_0\in\mathbb{C}$. To make this solution unique (up to the constant factor $C$) one have to require
\begin{equation}
\gamma(\sigma_0)\equiv <(\sigma_0+i\alpha W(s))^{-2}>\neq0, \label{vvedUslovie}
\end{equation}
where $<>$ denotes the averaging over the period. The analysis of \cite{BelFr} discovers the unstable eigenvalues under  assumption (\ref{vvedUslovie}). (In other words, the eigenvalue of the limit problem is required to be  simple.)  In particular,  this condition is satisfied for $W(s)=\sin s$ provided $\operatorname{Re}\sigma_0\neq0$. Then  $\sigma_0$
have to be selected using the solvability condition of the equation of  the next approximation to the eigenfunction. For a generic profile, however, $\gamma$ has zeroes outside the real axis. For such $\sigma_0$  Eq. (\ref{vvedH00}) has two independent solutions, i.e. $\sigma_0$ represents a multiple eigenvalue. In this paper, we develop the asymptotic in the case of multiple $\sigma_0$. In particular, we prove  that each zero of $\gamma$ generically gives rise to the unique branch of \emph{simple} eigenvalues $\sigma=\sigma(\varepsilon^2)$  while the  related eigenfunction has the form $\Phi(y)=f(my)$, i.e. $n$ must be equated to $0$ in  (\ref{Phin}), (\ref{H1}) and (\ref{H2}). As an example we examine the profile $W(s)=\sin s+\sin3s+\cos2s$. We calculate  the  concrete asymptotic expansion of the unstable eigenvalues and compare it with the numeric results. The comparison exhibits very good coincidence.

It should be noted  that function $\gamma(\sigma_0)$   has been introduced originally in \cite{Rosenbluth} in order to formulate  the necessary condition  for the long-wave  instability of the channel flows. However, the result of \cite{Rosenbluth}  exploits the monotonicity of the shear flow profile that makes it inapplicable to the oscillating  flows. At the same time, the asymptotic we found can be considered as long-wave one since  the problem (\ref{vved0}) in fact depends on $\alpha^2\varepsilon^2$ only when $n=0$.

\section*{Dispersion equation for an abstract spectral problem.}
Let $H=H(\sigma,\varepsilon)$ be linear operator defined for every $\sigma$ in some neighbourhood of $\sigma_0$ and for every  $\varepsilon$  in some  neighbourhood of zero. Assume that domains both of  $H$ and of $H^*$ do not depend on $\sigma$ and $\varepsilon$, the resolvent of $H$ is compact for every $\sigma$ and $\varepsilon$ and depends continuosly on $\sigma$ and $\varepsilon$ in uniform topology.
Consider the spectral problem
\begin{equation}
H(\sigma,\varepsilon)f=0, \label{Hpsi}
\end{equation}
where $\sigma$ is considered as spectral parameter. Assume that $H(\sigma_0,0){\buildrel\mathrm{def}\over=} H_{00}$ has non-trivial kernel: $\ker H_{00}\neq \{0\}$, i.~e. $\sigma_0$ is an eigenvalue for $\varepsilon=0$.  Let us reduce the
spectral problem \eqref{Hpsi} to a scalar equation for parameters $\sigma$ and $\varepsilon$.

Operator $H_{00}$ has compact resolvent, therefore its kernel is finite-dimen\-sional and $\dim \ker H_{00} = \dim \ker H_{00}^*$. Let $N=\dim \ker H_{00}$, $P_1$, $P_2$ -- projectors (may  be non self-adjoint) onto $N$-dimensional subspaces $\ker H_{00}$ and $\ker H_{00}^*$. Let $Q_k=I-P_k$, $k=1,2$. Set
\begin{equation}
f=P_1f+Q_1f{\buildrel\mathrm{def}\over=}\xi+\theta. \label{psixitheta}
\end{equation}
Apply $P_2$ and $Q_2$ to equation \eqref{Hpsi}. Write
\begin{equation}
P_2HP_1\xi+P_2HQ_1\theta=0, \label{abstrur1}
\end{equation}
\begin{equation}
Q_2HP_1\xi+Q_2HQ_1\theta=0. \label{abstrur2}
\end{equation}
Operator $Q_2H_{00}Q_1$ bijectively maps $\operatorname{Im} Q_1\cap {\cal D}(H)$ on $\operatorname{Im} Q_2$. Therefore the inverse operator $(Q_2H(\sigma,\varepsilon)Q_1)^{-1}:\operatorname{Im} Q_2\to\operatorname{Im} Q_1$ is bounded and continuous in a small neighborhood of $(\sigma_0,0)$.  We resolve  equation \eqref{abstrur2} in  $\theta$ and then substitute the solution into the \eqref{abstrur1}. This gives  an equation in a finite-dimensional subspace $\ker H_{00}$
$$
P_2H\xi-P_2HQ_1(Q_2HQ_1)^{-1}Q_2H\xi=0.
$$
We equate the determinant of this equation to zero and arrive to a scalar equation. This is the dispersion equation for  $\sigma$ and $\varepsilon$. The structure of the asymptotic of $\sigma(\varepsilon)$ can be deduced from Newton diagram (see \cite{VainTren}).

\section*{The limit operator kernel}
The operator $H_{00}=H(\sigma_0,0)$ of the problem \eqref{ishodnaya} is defined by
\begin{equation*}
\begin{split}
&H_{00}f=(\sigma_0+i\alpha W)f''-i\alpha W''f=\\
&=\frac{d}{d s}\left( (\sigma_0+i\alpha W)^2\frac d{d s}\frac f{\sigma+i\alpha W}  \right).
\end{split}
\end{equation*}
For all $\sigma_0$ it's kernel is non-trivial. It contains functions
\begin{equation}
f(s)=C(\sigma_0+i\alpha W) \label{reshvsegdaest}
\end{equation}
with arbitrary constant $C$. If equation
\begin{equation}
\frac{d}{d s}\left( (\sigma_0+i\alpha W)^2\frac d{d s}\frac f{\sigma+i\alpha W}  \right)=0 \label{kerH00}
\end{equation}
doesn't have solutions different from \eqref{reshvsegdaest} then one can apply the results \cite{BelFr}. We focus ourselves  on the case of 2-dimensional kernel of $H_{00}$ i.e. we assume that  equation \eqref{kerH00} has two linearly independent solutions.

Integrating equation \eqref{kerH00}, we obtain
$$
\frac d{d s}\frac f{\sigma+i\alpha W}=\frac{C_1}{(\sigma+i\alpha W)^2}.
$$
The solvability condition for this equation is
$$
C_1< (\sigma+i\alpha W)^{-2} > =0,
$$
where $<..>$ denotes average: $<f>=\frac1{2\pi}\int_{-\pi}^\pi f(s)\,ds$. Thus, the kernel of $H_{00}$ is 2-dimensional if and only if
\begin{equation}
\gamma(\sigma_0)\equiv < (\sigma+i\alpha W)^{-2} > =0. \label{uslovie2kr}
\end{equation}
Then the general solution of \eqref{kerH00} has the form
\begin{equation}
\begin{split}
\varphi_1&(s)=\sigma_0+i\alpha W(s),\quad \varphi_2(s)=\varphi_1\int\frac1{\varphi_1^2}=\\
&=(\sigma_0+i\alpha W(s)) \int_0^s\frac{dy}{(\sigma_0+i\alpha W(y))^2}.
\end{split}
\label{defphi12}
\end{equation}
 We note that  $\gamma(\sigma_0)$ is analytical function provided $\operatorname{Re} \sigma_0\neq0$. Its zeros are the  multiple eigenvalues we are looking for.

\section*{The kernel of the adjoint  operator}
Let $\sigma_0$ satisfy \eqref{uslovie2kr}. Consider  the adjoint operator
$$
H_{00}^*f=\frac1{\sigma_0^*-i\alpha W}\frac d{ds}\left((\sigma_0^*-i\alpha W)^2\frac{df}{ds}\right).
$$
Let us integrate equation
$$
\frac d{ds}\left((\sigma_0^*-i\alpha W)^2\frac{df}{ds}\right)=0,
$$
and then divide  the result by $(\sigma_0^*-i\alpha W)^2$, and then integrate  once more. As a result we have
$$
f=C_1+C_2\int\limits_0^s\frac{dy}{(\sigma_0^*-i\alpha W(y))^2}.
$$
According to \eqref{uslovie2kr},   $f$ is periodic for all $C_1$ and $C_2$. Therefore, functions
\begin{equation}
\psi_1(s)=1,\;\; \psi_2(s)=\!\!\int\!\!\frac1{(\varphi_1^*)^2}=\!\!\int\limits_0^s\!\!\frac{dy}{(\sigma_0^*-i\alpha W(y))^2}. \label{defpsi12}
\end{equation}
span  the  kernel of $H_{00}^*$.

\section*{Dispersion equation for the problem \eqref{ishodnaya}}
Let us construct dispersion equation for the case of 2-dimensional kernel of limit operator. Let $P_1$ and $P_2$ be the orthogonal projectors onto subspaces $\ker H_{00}$ and $\ker H_{00}^*$ . Functions $\psi_1$ and $\psi_2$ are orthogonal so that
$$
P_2f\!=\!\frac{(f,\psi_1)}{(\psi_1,\psi_1)}\psi_1+\frac{(f,\psi_2)}{(\psi_2,\psi_2)}\psi_2=<\!f\!>+\frac{(f,\psi_2)}{(\psi_2,\psi_2)}\psi_2.
$$
Although  $\varphi_1$ and $\varphi_2$ are not orthogonal, it is clear that projection $P_1f=0$  if and only if $(f,\varphi_1)=0$ and $(f,\varphi_2)=0$. We define projectors $Q_{1,2}$ setting $Q_{1,2}=I-P_{1,2}$.

Let  $\theta=Q_1f$. Then  into sum
\begin{equation}
\theta=\beta_1\theta^{(1)}+\beta_2\theta^{(2)}, \label{razlQ1f}
\end{equation}
where functions $\theta^{(1,2)}$ satisfy an equation
\begin{equation}
Q_2HQ_1\theta^{(i)}+Q_2H\varphi_i=0,\quad i=1,2. \label{urtheta12}
\end{equation}
(this representation can be seen  from \eqref{abstrur1}). Then the equation \eqref{abstrur2} in 2-dimensional subspace $\ker H_{00}^*$ is equivalent to \begin{equation*}
\begin{split}
&\beta_1\bigl(H(\varphi_1+\theta^{(1)}),\psi_1\bigr)+\beta_2\bigl(H(\varphi_2+\theta^{(2)}),\psi_1\bigr)=0,\\
&\beta_1\bigl(H(\varphi_1+\theta^{(1)}),\psi_2\bigr)+\beta_2\bigl(H(\varphi_2+\theta^{(2)}),\psi_2\bigr)=0.
\end{split}
\end{equation*}
The equating of the determinant of this system to zero gives us the dispersion equation
\begin{equation}
\Delta\equiv\begin{vmatrix}
\bigl(H(\varphi_1+\theta^{(1)}),\psi_1\bigr) & \bigl(H(\varphi_2+\theta^{(2)}),\psi_1\bigr)\\
\bigl(H(\varphi_1+\theta^{(1)}),\psi_2\bigr) & \bigl(H(\varphi_2+\theta^{(2)}),\psi_2\bigr)
\end{vmatrix}=0. \label{dispeq}
\end{equation}

\section*{The construction of the asymptotic}
Let's plot Newton diagram to determine the structure of the eigenvalue asymptotic (see \cite{VainTren}).
We expand function $\theta^{(1,2)}$, operator $H$ and determinant $\Delta$ in the powers of $\varepsilon$:
\begin{equation}
\theta^{(i)}=\theta^{(i)}_0+\varepsilon^1\theta^{(i)}_1+\varepsilon^2\theta^{(i)}_2+... \quad i=1,2; \label{razlthetaeps}
\end{equation}
\begin{equation}
H(\sigma,\varepsilon)=H_0(\sigma)+\varepsilon^1H_1(\sigma)+\varepsilon^2H_2(\sigma); \label{razlH012}
\end{equation}
\begin{equation}
\begin{split}
&H_0f\!=\!\phi f''\!-\phi''f\!=\!\!\left[ \phi^2 \left(\frac f\phi \right)'  \right]'\!\!;\;\; H_1f\!=\!2in\phi f';\\
&H_2f=-(\alpha^2+n^2)\phi f;\quad \phi(s)=\sigma+i\alpha W(s).
\end{split}\label{razlH}
\end{equation}
\begin{equation}
\Delta=\Delta_0+\varepsilon^1\Delta_1+\varepsilon^2\Delta_2+... \label{razlDelta}
\end{equation}
According to expansion \eqref{razlH012} equations \eqref{urtheta12} take the form
\begin{equation*}
\begin{split}
Q_2H_0\theta^{(i)}+&\varepsilon^1 Q_2H_1\theta^{(i)}+\varepsilon^2 Q_2H_2\theta^{(i)}=\\
 &=-Q_2H_0\varphi_i-\varepsilon^1 Q_2H_1\varphi_i - \varepsilon^2 Q_2H_2\varphi_i.
\end{split}
\end{equation*}
The substitution of \eqref{razlthetaeps}  gives series of equations for coefficients $\theta_k^{(i)}$, $i=1,2$
\begin{equation}
\begin{split}
& k\!=\!0: \;\; Q_2H_0\theta^{(i)}_0=-Q_2H_0\varphi_i,\\
& k\!=\!1: \;\; Q_2H_0\theta^{(i)}_1+Q_2H_1\theta^{(i)}_0=-Q_2H_1\varphi_i,\\
& k\!=\!2: \;\; Q_2H_0\theta^{(i)}_2+Q_2H_1\theta^{(i)}_1+Q_2H_2\theta^{(i)}_0=-Q_2H_2\varphi_i,\\
& k\!\geq\! 3: \;\; Q_2H_0\theta^{(i)}_k+Q_2H_1\theta^{(i)}_{k-1}+Q_2H_2\theta^{(i)}_{k-2}=0.
\end{split}\label{thetak}
\end{equation}

\textbf{Theorem 1.} \textit{
a) Coefficients $\Delta_0(\sigma)$ and $\Delta_1(\sigma)$ identically equal to zero.\\
b) If $n\neq 0$ then $\Delta_2(\sigma_0)\neq0$,  and such $\sigma_0$  produce no   analytical  branches  of   eigenvalues $\sigma(\varepsilon)$.   \\
c) If $n=0$ then $\Delta_2(\sigma_0)=0$ and  $\Delta_{2k+1}(\sigma)\equiv0$, $k\in\mathbb{N}$.\\
d) Let $n=0$ and  $\int_{-\pi}^\pi\varphi_1^{-3}\,ds\neq0$. If  $\sigma_0$ is real then assume additionally that $\sigma_0^2\neq\alpha^2<W^2>$. Then
$\sigma_0$ is a limit point for some branch of simple eigenvalues $\sigma(\varepsilon^2)$, and
\begin{equation}
\sigma=\sigma_0+\sigma_2\varepsilon^2+O(\varepsilon^4),\quad \sigma_2=-\frac{\Delta_4(\sigma_0)}{ \frac{d\Delta_2}{d\sigma}(\sigma_0) },\label{asymptotika}
\end{equation}
where $\Delta_4(\sigma_0)$ and $\frac{d\Delta_2}{d\sigma}(\sigma_0)$ are defined by equalities
$$
\Delta_4(\sigma_0)=\begin{vmatrix}
\bigl(H_2\varphi_1,\psi_1\bigr) & \bigl(H_2\varphi_2,\psi_1\bigr)\\
\bigl(H_2\varphi_1,\psi_2\bigr) & \bigl(H_2\varphi_2,\psi_2\bigr)
\end{vmatrix},
$$
$$
\frac{d\Delta_2}{d\sigma}(\sigma_0)=2\alpha^2<\varphi_1^2><\varphi_1^{-3}>.
$$}

To prove the theorem we have to expand  the scalar products $\bigl(H_i\varphi_j,\psi_k\bigr)$ and $\bigl(H_i\theta_m^{(j)},\psi_k\bigr)$ in the powers of  $\varepsilon$.

\textbf{Lemma 1.}
\textit{For all $\sigma$: $\operatorname{Re} \sigma\neq 0$ following statements hold:\\
a)~$\bigl(H_0\varphi_j,\psi_1\bigr)\!=\!\bigl(H_0\theta_m^{(j)},\psi_1\bigr)\!=\!0$;  \\
b)~$\bigl(H_1\varphi_1,\psi_1\bigr)=0$; \hskip5mm
c)~$\bigl(H_1\varphi_2,\psi_1\bigr)=in$; \\
d)~$\bigl(H_1\varphi_1,\psi_2\bigr)=-in$; \hskip5mm
e)~$\bigl(H_0\varphi_1,\psi_2\bigr)=0$ \\
f)~If $\sigma\neq\sigma_0$ then any solution of $Q_2H_0\xi=0$ can be represented in the form $\xi=C_1\phi+C_2\widetilde\xi$, where $\widetilde\xi$~-- particular solution of $H_0\widetilde\xi=\psi_2$, function $\phi$ defined in \eqref{razlH}.
}

{\bf Proof of the lemma.} Equalities a,b,c,d and e are straightforward. To get statement~f one can use the equivalence of equations $Q_2H_0\xi=0$ and $H_0\xi=\alpha_1\psi_2+\alpha_2\psi_2$. Then $\alpha_1=0$ (according to the solvability condition). Then the general solution of homogeneous equation $H_0\xi=0$ is $C_1\phi$ while $\alpha_2\widetilde\xi$ is particular solution of the equation $H_0\xi=\alpha_2\psi_2$. This completes the proof.
\vskip1mm

{\bf Proof of the theorem 1.}

{\bf a)} According to statement a of lemma 1
\begin{equation*}
\Delta_0\!=\!\begin{vmatrix}
\bigl(H_0(\varphi_1+\theta^{(1)}_0),\psi_1\bigr) & \bigl(H_0(\varphi_2+\theta^{(2)}_0),\psi_1\bigr)\\
\bigl(H_0(\varphi_1+\theta^{(1)}_0),\psi_2\bigr) & \bigl(H_0(\varphi_2+\theta^{(2)}_0),\psi_2\bigr)
\end{vmatrix}\!=\!0.
\end{equation*}
$$
\Delta_1=\begin{vmatrix}
\bigl(H_1(\varphi_1+\theta^{(1)}_0),\psi_1\bigr) & \bigl(H_1(\varphi_2+\theta^{(2)}_0),\psi_1\bigr)\\
\bigl(H_0(\varphi_1+\theta^{(1)}_0),\psi_2\bigr) & \bigl(H_0(\varphi_2+\theta^{(2)}_0),\psi_2\bigr)
\end{vmatrix} +
$$
$$
+ \begin{vmatrix}
\bigl(H_0\theta^{(1)}_1,\psi_1\bigr) & \bigl(H_0\theta^{(2)}_1,\psi_1\bigr)\\
\bigl(H_0(\varphi_1+\theta^{(1)}_0),\psi_2\bigr) & \bigl(H_0(\varphi_2+\theta^{(2)}_0),\psi_2\bigr)
\end{vmatrix} +
$$
$$
+\begin{vmatrix}
\bigl(H_0(\varphi_1+\theta^{(1)}_0),\psi_1\bigr) & \bigl(H_0(\varphi_2+\theta^{(2)}_0),\psi_1\bigr)\\
\bigl(H_1(\varphi_1+\theta^{(1)}_0),\psi_2\bigr) & \bigl(H_1(\varphi_2+\theta^{(2)}_0),\psi_2\bigr)
\end{vmatrix}+
$$
$$
+
\begin{vmatrix}
\bigl(H_0(\varphi_1+\theta^{(1)}_0),\psi_1\bigr) & \bigl(H_0(\varphi_2+\theta^{(2)}_0),\psi_1\bigr)\\
\bigl(H_0\theta^{(1)}_1,\psi_2\bigr) & \bigl(H_0\theta^{(2)}_1,\psi_2\bigr)
\end{vmatrix}=$$
$$
=
\begin{vmatrix}
\bigl(H_1(\varphi_1+\theta^{(1)}_0),\psi_1\bigr) & \bigl(H_1(\varphi_2+\theta^{(2)}_0),\psi_1\bigr)\\
\bigl(H_0(\varphi_1+\theta^{(1)}_0),\psi_2\bigr) & \bigl(H_0(\varphi_2+\theta^{(2)}_0),\psi_2\bigr)
\end{vmatrix}.
$$
According to statement f of the lemma $\varphi_m+\theta_0^{(m)}=C_1^{(m)}\phi+C_2^{(m)}\widetilde\xi$. Then
$$
\bigl(H_0(\varphi_m+\theta^{(m)}_0),\psi_2\bigr)\!=\!C_2^{(m)}\bigl(H_0\widetilde\xi,\psi_2\bigr)\!=\!C_2^{(m)}\bigl(\psi_2,\psi_2\bigr),
$$
$$
\bigl(H_1(\varphi_m+\theta^{(m)}_0),\psi_1\bigr)=C_2^{(m)}\bigl(H_1\widetilde\xi,\psi_1\bigr).
$$
(the statement b of lemma~1 was used for the last equality). Thus, we get a determinant with proportional rows
$$
\Delta_1=\begin{vmatrix}
C_2^{(1)}\bigl(H_1\widetilde\xi,\psi_1\bigr) & C_2^{(2)}\bigl(H_1\widetilde\xi,\psi_1\bigr)\\
C_2^{(1)}\bigl(\psi_2,\psi_2\bigr) & C_2^{(2)}\bigl(\psi_2,\psi_2\bigr)
\end{vmatrix}=0.
$$

{\bf b)} Writing coefficient $\Delta_2$ similarly to $\Delta_1$ we obtain the sum of eight determinants. Half of them are equal to zero according to the statement a of lemma~1. As a result we arrive at  equality
\begin{equation}
\begin{split}
\Delta_2&=\begin{vmatrix}
\bigl(H_2(\varphi_1+\theta^{(1)}_0),\psi_1\bigr) & \bigl(H_2(\varphi_2+\theta^{(2)}_0),\psi_1\bigr)\\
\bigl(H_0(\varphi_1+\theta^{(1)}_0),\psi_2\bigr) & \bigl(H_0(\varphi_2+\theta^{(2)}_0),\psi_2\bigr)
\end{vmatrix}+\\
+&\begin{vmatrix}
\bigl(H_1\theta^{(1)}_1,\psi_1\bigr) & \bigl(H_1\theta^{(2)}_1,\psi_1\bigr)\\
\bigl(H_0(\varphi_1+\theta^{(1)}_0),\psi_2\bigr) & \bigl(H_0(\varphi_2+\theta^{(2)}_0),\psi_2\bigr)
\end{vmatrix}+\\
&+\begin{vmatrix}
\bigl(H_1(\varphi_1+\theta^{(1)}_0),\psi_1\bigr) & \bigl(H_1(\varphi_2+\theta^{(2)}_0),\psi_1\bigr)\\
\bigl(H_1(\varphi_1+\theta^{(1)}_0),\psi_2\bigr) & \bigl(H_1(\varphi_2+\theta^{(2)}_0),\psi_2\bigr)
\end{vmatrix}+\\
+&\begin{vmatrix}
\bigl(H_1(\varphi_1+\theta^{(1)}_0),\psi_1\bigr) & \bigl(H_1(\varphi_2+\theta^{(2)}_0),\psi_1\bigr)\\
\bigl(H_0\theta^{(1)}_1,\psi_2\bigr) & \bigl(H_0\theta^{(2)}_1,\psi_2\bigr)
\end{vmatrix}.
\end{split}\label{Delta2}
\end{equation}
Let $\sigma=\sigma_0$. Then the first, second and forth determinants are equal to zero as  $\operatorname{Im} H_{00} \perp \psi_{1,2}$.
since  $H_{00}\varphi_{1,2}=0$ equations \eqref{thetak} yield $\theta_0^{(1,2)}=0$ provided $\sigma=\sigma_0$. Using statements b,c and d of the lemma~1 we get
$$
\Delta_2(\sigma_0)=\begin{vmatrix}
0 & in\\
-in & \bigl(H_1\varphi_2,\psi_2\bigr)
\end{vmatrix}=-n^2.
$$
\begin{center}
\includegraphics[width=70mm, height=40mm]{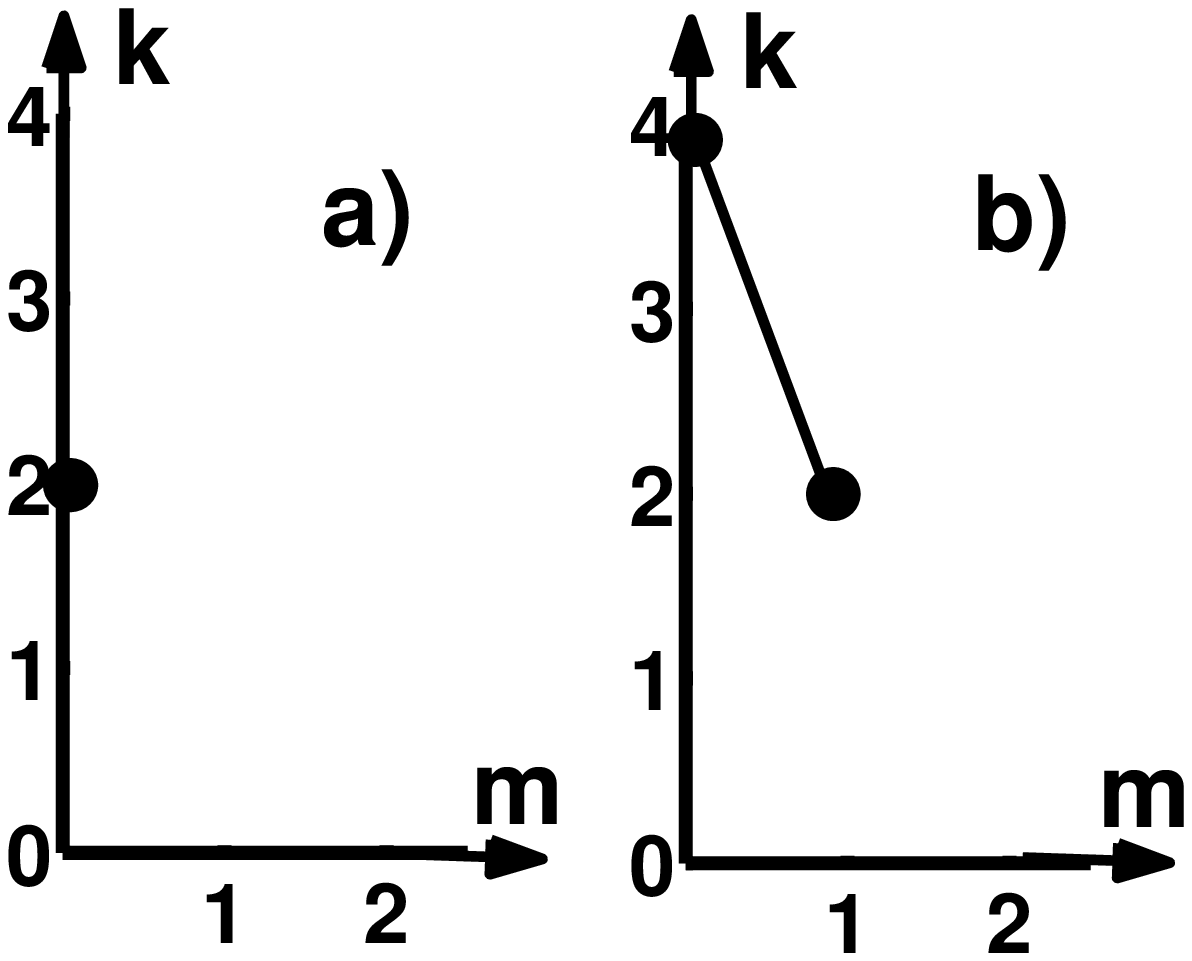}\\
Fig.~1. Newton diagram
\end{center}
\vskip-2mm If $n\neq0$ then $\Delta_2(\sigma_0)\neq0$. The related  Newton diagram is presented on fig.~1a. Line $k=0$ corresponds to the coefficient $\Delta_0$, line $k=1$~--- to the coefficient $\Delta_1$  and so on. Since $\Delta_2(\sigma_0)\neq0$, Newton diagram includes point $(0,2)$, while there are no points below the line $k=2$ (due to statement a of theorem~1). Therefore, equation \eqref{dispeq} doesn't have solutions nearby $\sigma_0$ when $\varepsilon\to 0$.

{\bf c)} If $n=0$ then $H_1=0$ and operator $H$ in fact depends on  $\varepsilon^2$. Consequently, coefficients $\Delta_{2k+1}$  are equal to zero.

Let's calculate $\Delta_4(\sigma_0)$. Taking into account that  $\theta_0^{(1,2)}(\sigma_0)= 0$  and  $\operatorname{Im} H_{00} \perp \psi_{1,2}$, we get
$$
\Delta_4(\sigma_0)=\begin{vmatrix}
\bigl(H_2\varphi_1,\psi_1\bigr) & \bigl(H_2\varphi_2,\psi_1\bigr)\\
\bigl(H_2\varphi_1,\psi_2\bigr) & \bigl(H_2\varphi_2,\psi_2\bigr)
\end{vmatrix}.
$$
Let's calculate $\frac{d\Delta_2}{d\sigma}(\sigma_0)$. Taking into account that $H_1=0$ and  \eqref{Delta2} we get
$$
\frac{d\Delta_2}{d\sigma}(\sigma_0)=\begin{vmatrix}
\bigl(H_2\varphi_1,\psi_1\bigr) & \bigl(H_2\varphi_2,\psi_1\bigr)\\
\bigl(\frac{\partial H_0}{\partial \sigma}\varphi_1,\psi_2\bigr) & \bigl(\frac{\partial H_0}{\partial \sigma}\varphi_2,\psi_2\bigr)
\end{vmatrix}.
$$
 Then $\bigl(\frac{\partial H_0}{\partial \sigma}\varphi_1,\psi_2\bigr)=0$ by statement e of the lemma. Since $\frac{\partial H_0}{\partial \sigma}f=f''$, we have
\begin{equation}
\frac{d\Delta_2}{d\sigma}(\sigma_0)=
\bigl(H_2\varphi_1,\psi_1\bigr) \bigl(\varphi_2'',\psi_2\bigr). \label{dDelta2ds}
\end{equation}

Let's determine scalar products $\bigl(H_2\varphi_1,\psi_1\bigr)$ and $\bigl(\varphi_2'',\psi_2\bigr)$.
$$
\bigl(H_2\varphi_1,\psi_1\bigr)=-\frac{\alpha^2}{2\pi}\int_{-\pi}^\pi \varphi_1^2\,ds=-\alpha^2\bigl[\operatorname{Re}(\sigma^2)-\alpha^2 <W^2>+ i\sigma_{0_{re}}\sigma_{0_{im}}\bigr].
$$
If $\sigma_{0_{im}}\neq0$, then imaginary part of the scalar product is nonzero otherwise  we have to require that $\sigma_0^2\neq\alpha^2<W^2>$ in order ensure that $\frac{d\Delta_2}{d\sigma}(\sigma_0)$ is nonzero. Let's transform the second multiplier in formula~\eqref{dDelta2ds}
\begin{equation*}
\begin{split}
&\bigl(\varphi_2'',\psi_2\bigr)\!=\!\frac{1}{2\pi}\!\int\limits_{-\pi}^\pi \!\!\varphi_2 \Bigl(\frac1{\varphi_1^2}\Bigr)'ds\!=\! \frac{1}{\pi}\!\int\limits_{-\pi}^\pi \!\!\varphi_1 \Bigl(\int\!\!\frac1{\varphi_1^2}\Bigr)\frac{-\varphi_1'}{\varphi_1^3}ds\!=\\
&=\frac{1}{\pi}\int\limits_{-\pi}^\pi \Bigl(\frac{1}{\varphi_1}\Bigr)' \Bigl(\int\frac1{\varphi_1^2}\Bigr) \,ds=
-\frac{1}{\pi}\int\limits_{-\pi}^\pi \frac1{\varphi_1^3} \,ds.
\end{split}
\end{equation*}
Assuming that  $<\varphi_1^{-3}>\neq0$ we arrive at the  Newton diagram presented on fig.\,1b which, in turn, implies  \eqref{asymptotika}. The proof is completed.

\section*{Computing experiment}
Consider flow  profile $W(s)=\sin s+\sin3s+\cos2s$. If $n=0$, then problem \eqref{ishodnaya} depends on only one parameter: after substitution $\sigma=i\alpha \widetilde \sigma$, $\varepsilon=\widetilde\varepsilon/\alpha$ only one parameter $\widetilde\varepsilon$  and unknown eigenvalue $\widetilde\sigma$ left. Therefore, we can set $\alpha=1$ with no losses in  generality. Function $\gamma$ defined in \eqref{uslovie2kr} vanishes in the point $\sigma_0=0.3543-0.6366i$. Coefficient $\sigma_2$ of asymptotic expansion \eqref{asymptotika} is $\sigma_2=-0.2568 + 0.2393i$. The  asymptotic
$
\sigma_{ap}=\sigma_0+\sigma_2\varepsilon^2+\ldots
$
demonstrates very good agreement with the Numerical results (see fig.~2 and 3).

\begin{center}
\vskip-3mm
\includegraphics[width=70mm, height=40mm]{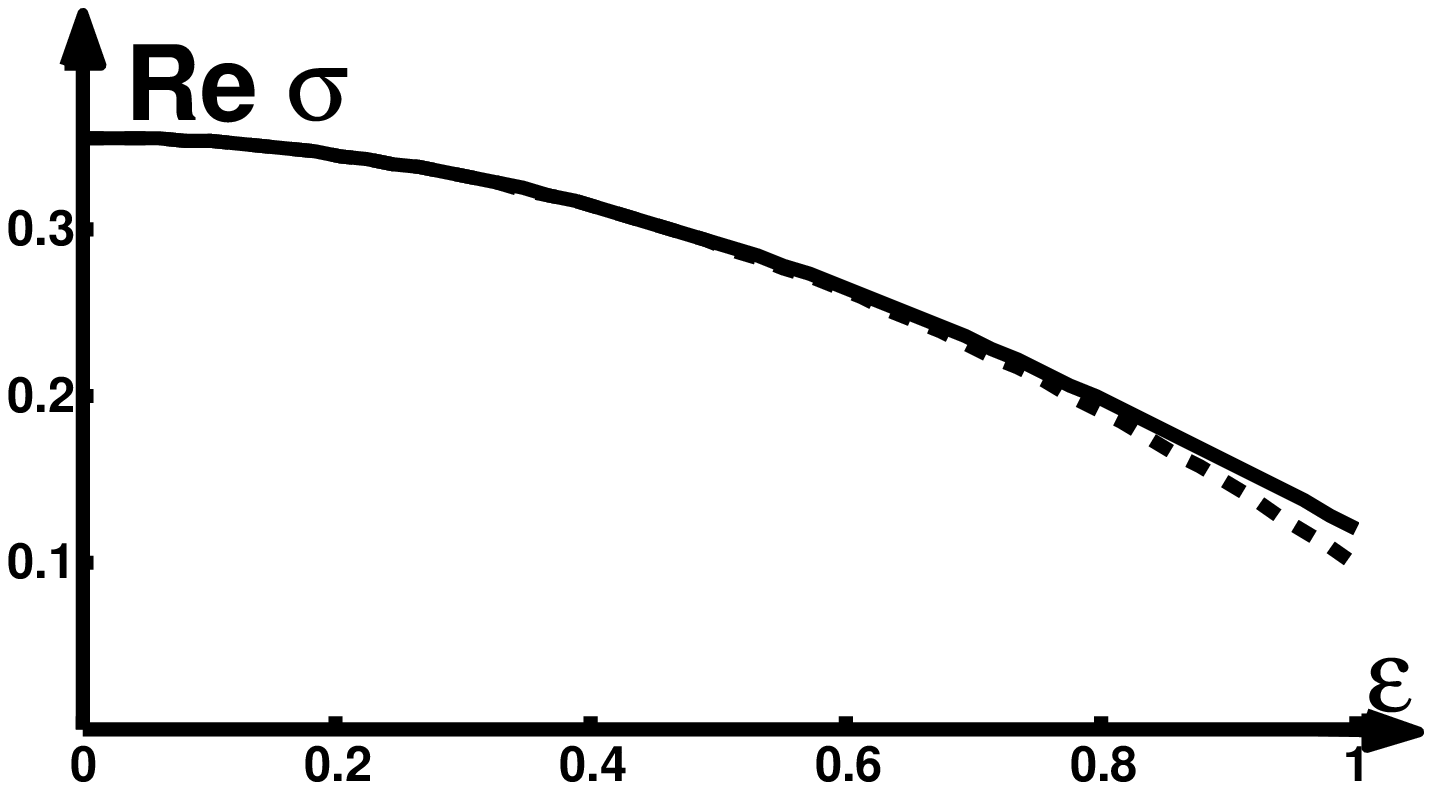}\\  \vskip-1mm
\includegraphics[width=70mm, height=40mm]{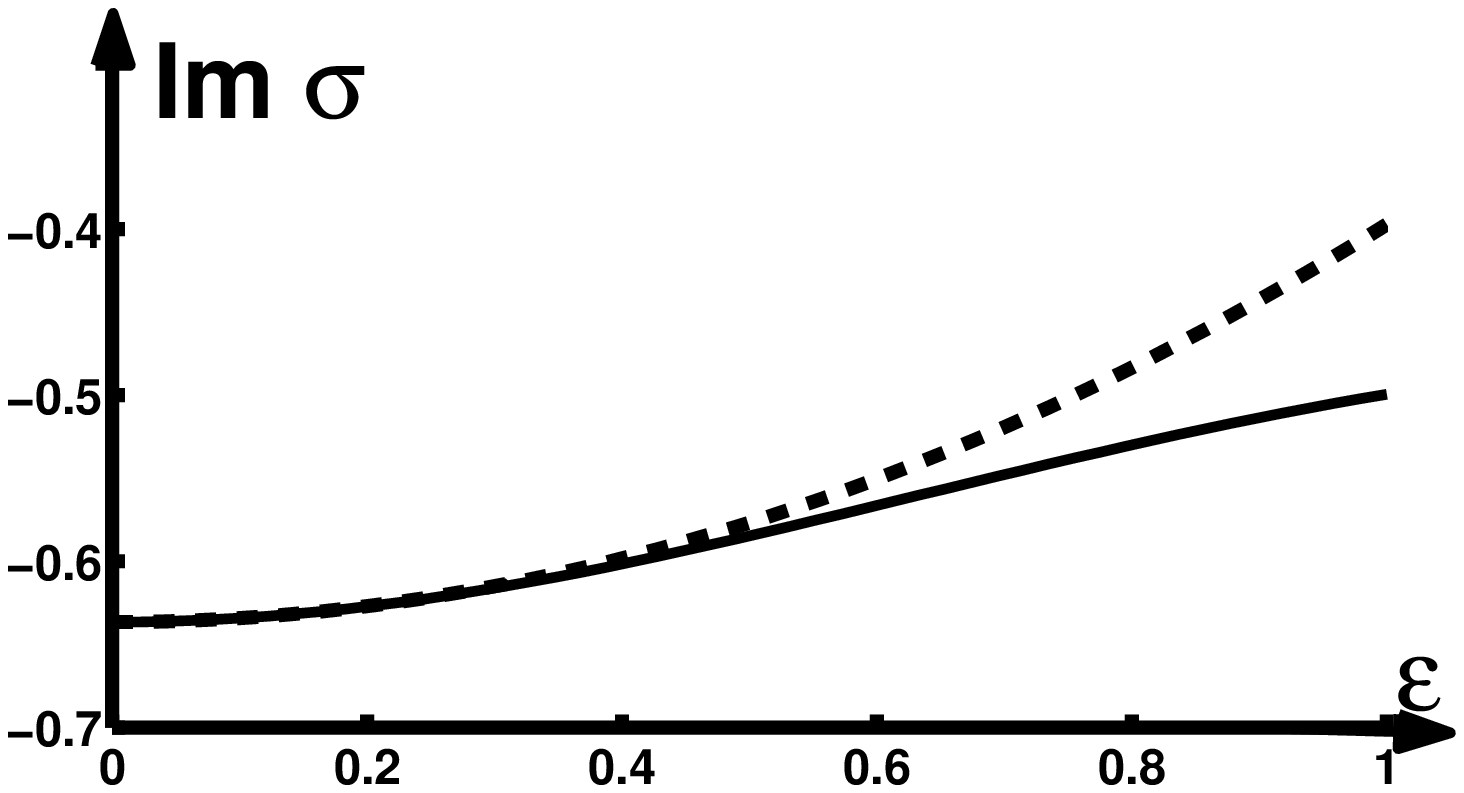}\\  \vskip-1mm
{Fig.~2. Plots of the numerical (solid line)  and asymptotic (dashed line) solution $\sigma(\varepsilon)$. Abscissa is~$\varepsilon$, ordinate axises are $\operatorname{Re}\sigma$ and $\operatorname{Im}\sigma$ respectively.}
\end{center}

\begin{center}
\vskip-3mm
\includegraphics[width=70mm, height=50mm]{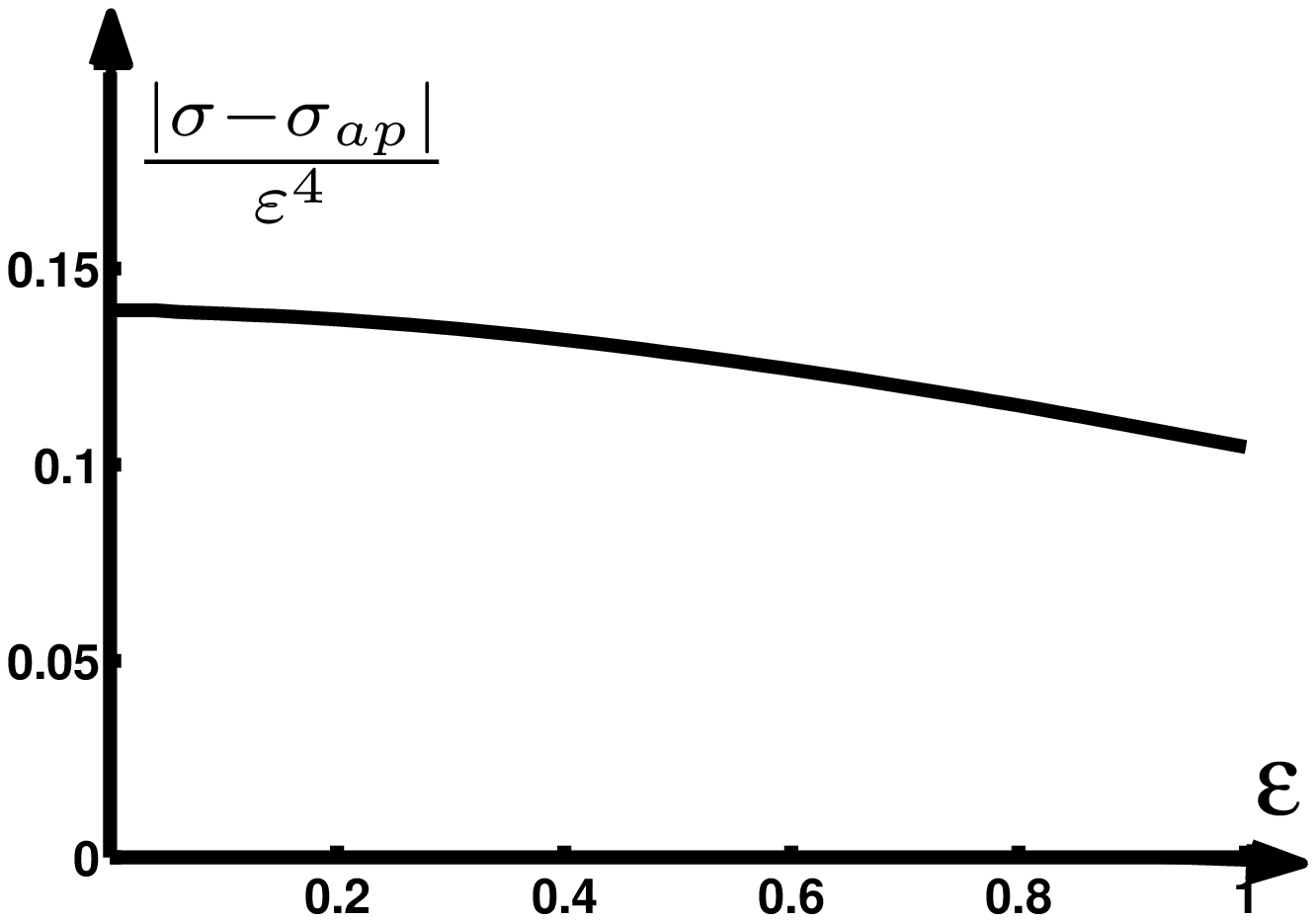}\\  \vskip-1mm
{Fig.~3. The relative difference between the numerical solution and  asymptotic. Abscissa is~$\varepsilon$, ordinate axis is $\frac{|\sigma-\sigma_{ap}|}{\varepsilon^4}$.}
\end{center}

\vskip-2mm
{\bf Acknowledgements.} I thank Andrey Morgulis for the problem definition, fertile conversations and great help in translation of the article. This research was performed with the support of the U.S. Civilian Research and Development Foundation (grant RUM1-2842-RO-06) and Russian Foundation for Fundamental Research (grant 08-01-00895-a).

\end{document}